\documentclass[]{aastex631}
\usepackage{longtable}
\usepackage{footnote}
\usepackage{booktabs}
\usepackage{amsmath}
\usepackage[figuresright]{rotating}
\usepackage{threeparttable}
\usepackage{appendix}

\graphicspath{{./}{figures/}}

\begin{document}
\correspondingauthor{Qian Sheng-Bang}
\email{qiansb@ynu.ac.cn}

\title{A brown dwarf orbiting around the planetary-nebula central binary KV Vel}

\author{Qian S.-B.}
\affiliation{Department of Astronomy, School of Physics and Astronomy, Yunnan University, 650091 Kunming, China}
\affiliation{Key Laboratory of Astroparticle Physics of Yunnan Province, Yunnan University, 650091 Kunming, China}

\author{Zhu L.-Y.}
\affiliation{Yunnan Observatories, Chinese Academy of Sciences (CAS), P.O. Box 110, 650216 Kunming, China}
\affiliation{University of Chinese Academy of Sciences, No.19(A) Yuquan Road, Shijingshan District, 100049 Beijing, China}

\author{Li F.-X.}
\affiliation{Department of Astronomy, School of Physics and Astronomy, Yunnan University, 650091 Kunming, China}
\affiliation{Key Laboratory of Astroparticle Physics of Yunnan Province, Yunnan University, 650091 Kunming, China}

\author{Li L.-J.}
\affiliation{Yunnan Observatories, Chinese Academy of Sciences (CAS), P.O. Box 110, 650216 Kunming, China}

\author{Han Z.-T.}
\affiliation{Department of Physics, Yuxi Normal University, Phoenix Road 134, 653100 Yuxi, China}

\author{He J.-J.}
\affiliation{Yunnan Observatories, Chinese Academy of Sciences (CAS), P.O. Box 110, 650216 Kunming, China}

\author{Zang L.}
\affiliation{Department of Astronomy, School of Physics and Astronomy, Yunnan University, 650091 Kunming, China}
\affiliation{Key Laboratory of Astroparticle Physics of Yunnan Province, Yunnan University, 650091 Kunming, China}

\author{Chang L.-F.}
\affiliation{Department of Astronomy, School of Physics and Astronomy, Yunnan University, 650091 Kunming, China}
\affiliation{Key Laboratory of Astroparticle Physics of Yunnan Province, Yunnan University, 650091 Kunming, China}

\author{Sun Q.-B.}
\affiliation{Department of Astronomy, School of Physics and Astronomy, Yunnan University, 650091 Kunming, China}
\affiliation{Key Laboratory of Astroparticle Physics of Yunnan Province, Yunnan University, 650091 Kunming, China}

\author{Li M.-Y.}
\affiliation{Department of Astronomy, School of Physics and Astronomy, Yunnan University, 650091 Kunming, China}
\affiliation{Key Laboratory of Astroparticle Physics of Yunnan Province, Yunnan University, 650091 Kunming, China}

\author{Zhang H.-T.}
\affiliation{Department of Astronomy, School of Physics and Astronomy, Yunnan University, 650091 Kunming, China}
\affiliation{Key Laboratory of Astroparticle Physics of Yunnan Province, Yunnan University, 650091 Kunming, China}

\author{Yan F.-Z.}
\affiliation{Department of Astronomy, School of Physics and Astronomy, Yunnan University, 650091 Kunming, China}
\affiliation{Key Laboratory of Astroparticle Physics of Yunnan Province, Yunnan University, 650091 Kunming, China}

\begin{abstract}
KV Vel is a non-eclipsing short-period (P= 0.3571\,days) close binary containing a
very hot subdwarf primary (77000\,K) and a cool low-mass secondary star (3400\,K) that is located at the center of the planetary nebula DS 1.
The changes in the orbital period of the close binary were analyzed based on 262 new times of light maximum together with those compiled from the literature.
It is discovered that the O-C curve shows
a small-amplitude ($0.^{d}0034$) cyclic period variation with a period of 29.55\,years.
The explanation by the solar-type magnetic activity cycles of the cool component is ruled out because the required energies are much
larger than the total radiant energy of this component in a whole cycle. Therefore, the cyclic
variation was plausibly explained as the light-travel time effect via
the presence of a tertiary component, which is supported by the periodic changes of the O-C curve and the rather symmetric and stable light curves obtained by TESS.
The mass of the tertiary companion is determined to be
$M_3\sin{i^{\prime}}=0.060(\pm0.007)$\,$M_{\odot}$.
If the third body is coplanar with the central binary (i.e., $i^{\prime}={62.5}^{\circ}$),
the mass of the tertiary component is computed as $M_3\sim0.068$\,$M_{\odot}$, and thus it would be below the stable hydrogen-burning limit and is a brown dwarf.
The orbital separation is shorter than 9.35 astronomical units (AU).
KV Vel together with its surrounding planetary nebula and the brown-dwarf companion may be formed through
the common-envelope evolution after the primary filled its Roche lobe during the early asymptotic giant branch stage.

\end{abstract}

\keywords{
Stars: binaries : close --
          Stars: binaries : eclipsing --
          Stars: individuals (KV Vel) --
          Stars: subdwarfs --
          Stars: low-mass, brown dwarfs
}

\section{Introduction} \label{sec:intro}
About one in five planetary nebulae (PNe) are formed from the evolutions of common envelopes
(e.g., \citealt{2009A&A...505..249M}). The central stars of these planetary nebulae are usually close binaries with a very hot
subdwarf star and a very cool secondary star in detached configurations. Therefore, they usually show very strong reflection effects with amplitudes exceeding one
magnitude and emission line dominated spectra (e.g., \citealt{2006A&A...456.1069S}). Most of them are likely evolving from the asymptotic giant branch (AGB) (e.g., \citealt{1993ApJ...418..343I}).
The embedded planetary nebulae in these binary systems suggest that they are very young from the common-envelopes evolution. They are important targets to investigate the evolution of common envelopes and are the progenitors of cataclysmic variables. Some exoplanets or brown dwarfs have been found to be orbiting around subdwarf and white-dwarf binaries (e.g., \citealt{2009ApJ...695L.163Q,2012ApJ...745L..23Q,2013MNRAS.436.1408Q,2012A&A...543A.138B,2018ApJ...868...53H,2019RAA....19..134Z,2020MNRAS.499.3071S,2022RNAAS...6...94C}). Among of them, V471 Tau (DAZ+K2V; P = 12.52 hr) is very special for understanding fundamental aspects of stellar astrophysics and
binary evolution \citep{2022RNAAS...6...94C}, which is a member of the Hyades cluster and a brown-dwarf companion is detected (e.g., \citealt{2022RNAAS...6...94C,2022MNRAS.517.5358K}). However, no substellar objects companion to the central binaries of planetary nebulae have been reported to date.

KV Vel was originally discovered as an sdO star that possesses a planetary nebula DS 1 \citep{1983ApJ...270L..13D,1978A&AS...31...15H}. The nebula is
roughly circular with a diameter of $\sim$ $180^{\prime \prime}$. Later, radial-velocity curves for both binary components were measured by \citet{1985ApJ...294L.107D} with an orbital period of 8.57 hours,
indicating that it is the first planetary-nebula central star discovered to be a
double-lined spectroscopic binary. The binary system consists of a hot subdwarf and a cool M dwarf with a temperature difference up to 73600\,K \citep{1996MNRAS.279.1380H,2011A&A...526A.150R}. Photometric light curves have been published by \citet{1985ApJ...294L.107D}, \citet{1986AJ.....91.1372L}, \citet{1988Obs...108...88K}, and \citet{2011A&A...526A.150R}, which show
a very strong reflection effect of the secondary star with an amplitude of $\Delta{V}\sim0.55$ mag. in V \citep{1996MNRAS.279.1380H} and $\Delta{H}\sim0.7$ at near-infrared
wavelengths. It is one of the most pronounced such effects observed to date.  The symmetric shape
of the reflection effect with respect to the phase of maximum indicates that it is produced by uniform irradiation of the cool component by a source centered on the position of the extremely hot subdwarf primary. The very large-amplitude and the symmetric shape of reflection effect indicates that the times of maximum can be determined in high precision. Therefore, the changes in the orbital period could be investigated in details.

\section{Variations in the orbital period change} 

The orbital period of KV Vel, the central binary star surrounded by
a planetary nebula DS 1 \citep{1983ApJ...270L..13D,1978A&AS...31...15H}, was first determined as 8.571 hours by \citet{1985ApJ...294L.107D}. Based on multi-colour photometry observations,
\citet{1986AJ.....91.1372L} gave the first photometric ephemeris,
\begin{equation}
T_{max}= JD\,2445834.5268(9) + 0.357113(2) \times{E}.
\end{equation}
This ephemeris was later improved by \citet{1988Obs...108...88K} by using Drilling data and
their own data as,
\begin{equation}
T_{max} = HJD\,244 5834.5269(6) + 0.35711296(30) \times{E}.
\end{equation}
Near-infrared light curves in JHK bands were obtained by \citet{2011A&A...526A.150R}. They determined one time of maximum
light and found that the maximum time is displaced by 0.01 phases
from the ephemeris of \citet{1988Obs...108...88K}, which cannot be accounted for
by the uncertainty. Therefore, a revised photometric ephemeris,
\begin{equation}
T_{max} = HJD\,2445834.5174(4)+ 0.3571205(5)\times{E},
\end{equation}
was derived by \citet{2011A&A...526A.150R} by adding the near-infrared time of maximum.

\begin{figure}
\begin{center}
\includegraphics[angle=0,scale=0.3]{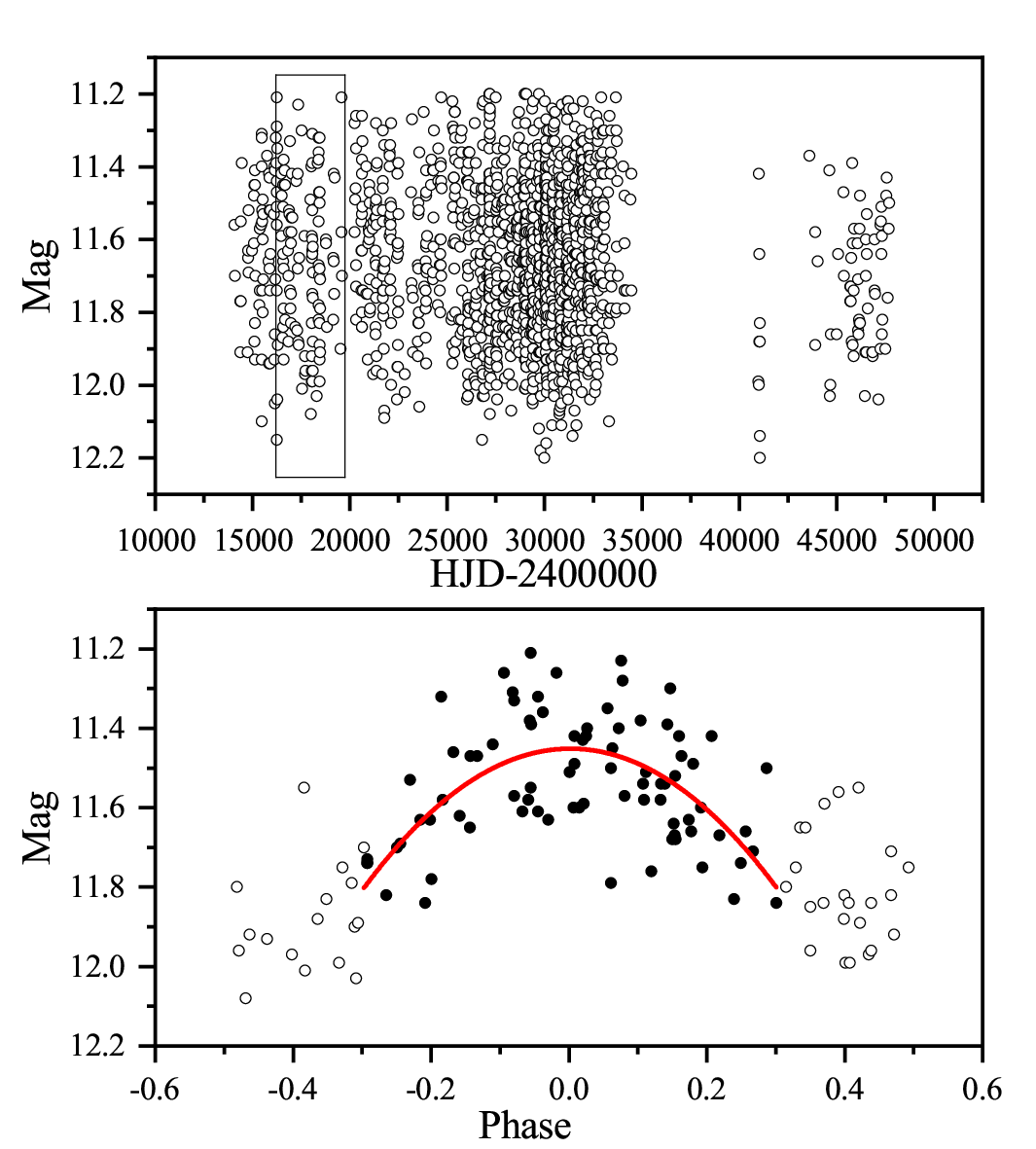}
\includegraphics[angle=0,scale=0.3]{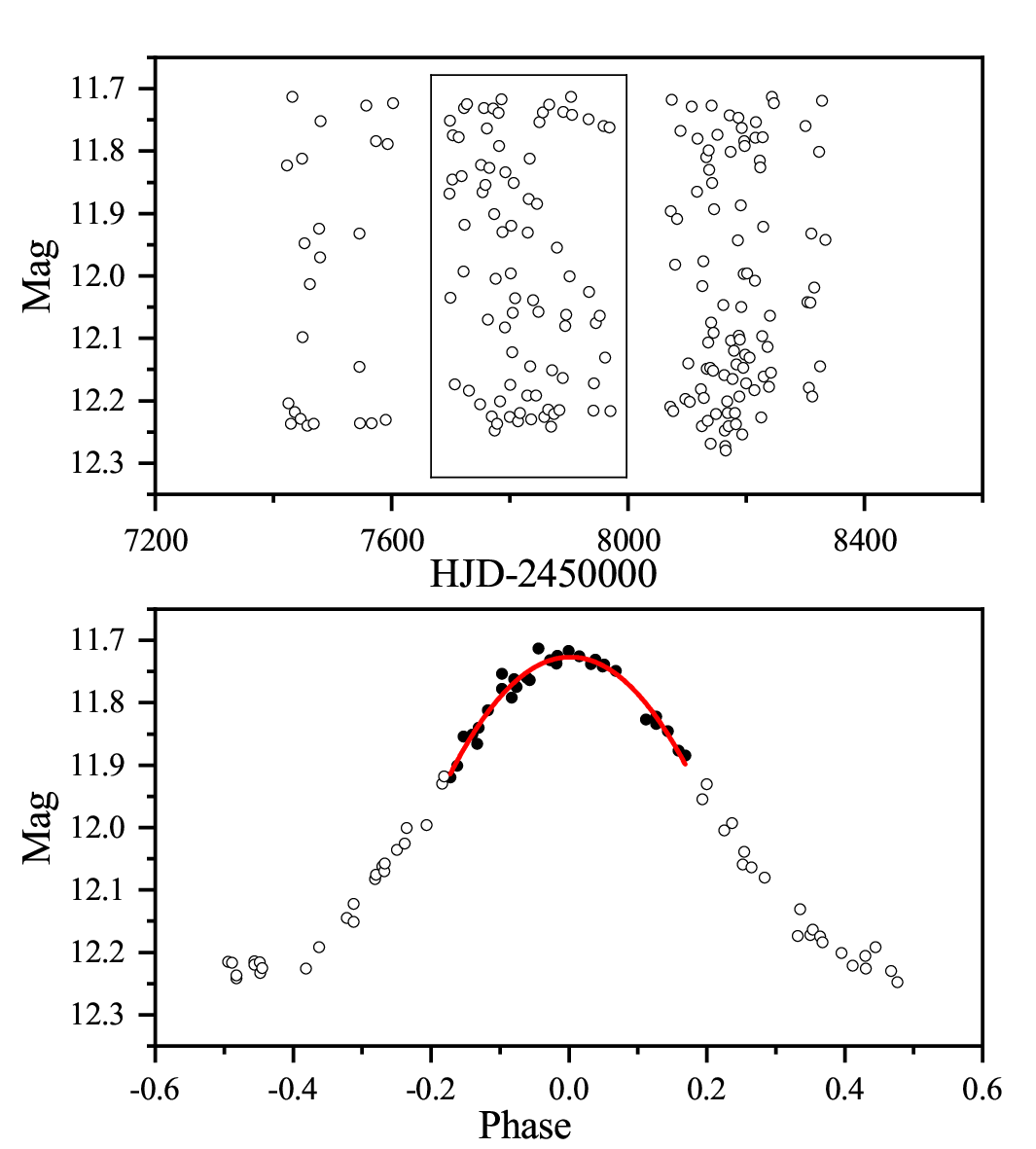}
\includegraphics[angle=0,scale=0.4]{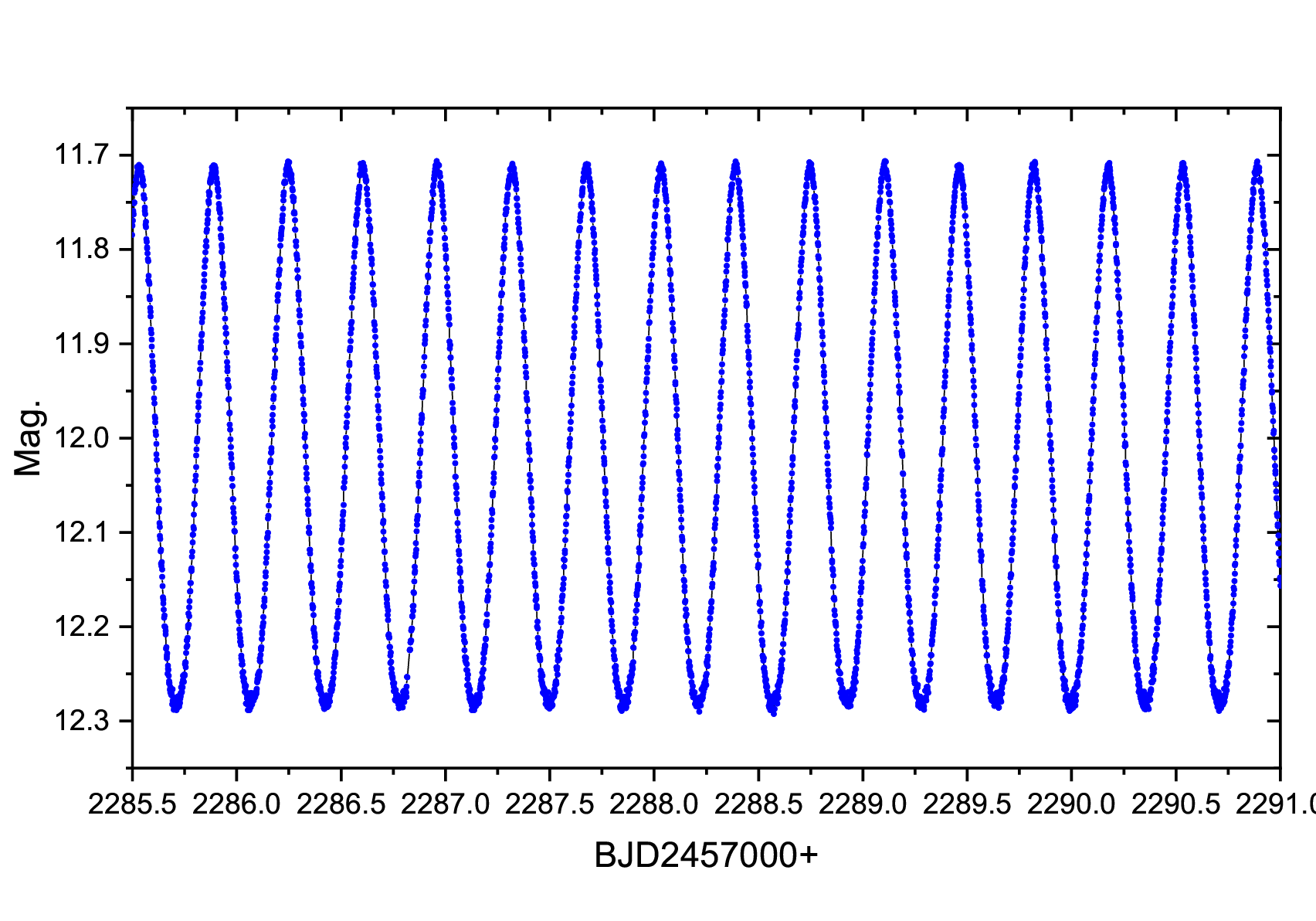}
\caption{Upper panels: Maxima of KV Vel from DASCH (left panel) and ASAS-SN (right panel). In each case, the upper panel shows the times-series photometric data and the lower panel displays the orbital phased light curve with the fitted parabola around the time of maximum. The filled circles in the two panels refer to the data around the maxima and are used to calculate the times of light maximum, while the open circles to those not used for time determining.} Lower panel: the part of the light curves obtained by TESS. \label{fig:FIT}
\end{center}

\end{figure}

Long-term photometric observations are very important to investigate the change in the orbital period of close binary stars.
The Digital Access to a Sky Century at Harvard project (DASCH) is committed to digitize the glass photographic plates that have accumulated for more than one hundred years, and provide photometric data available on line \citep{2009ASPC..410..101G,2012IAUS..285...29G}. As for KV Vel, there are 1567 DASCH observations in the time interval between HJD\,2414077 and HJD\,2447708. 21 times of light maximum were determined from the DASCH data by using a method similar to the AFP method \citep{2014A&A...572A..71Z}, which was also demonstrated and adopted by us (e.g., \citealt{2015AJ....149..148L,2023ApJ...956...49L}). By using the same method, several times were also computed by using the ASAS-SN database \citep{2014ApJ...788...48S,2019MNRAS.486.1907J}.
Two of the examples are shown in upper panels of Fig. \ref{fig:FIT}, where one time of light maximum in the left panel was determined from DASCH in the time interval between HJD 2416000 and 2419500, while the other one derived from the ASAS-SN data is displayed in the right panel. The uncertainties associated with the times of light maximum in the light curve were determined through the calculation of the covariance. Moreover, KV Vel was observed by TESS \citep{2015JATIS...1a4003R} in four sectors from April 2019 to March 2023, i.e., S10, S36, S37, and S63. A part of the TESS light curves is plotted in the lower panel of Fig. \ref{fig:FIT}. As shown in the panel, the TESS light curve is symmetric and rather stable. A large number of times of maximum light were obtained with TESS data. In total, 262 new times of light maximum have been determined and the time span of our data is more than 120 years (HJD 2416120 - 260040.5), which is very useful to analyze the long-term changes in the orbital period of KV Vel.

\begin{figure}
\begin{center}
\includegraphics[angle=0,scale=0.55]{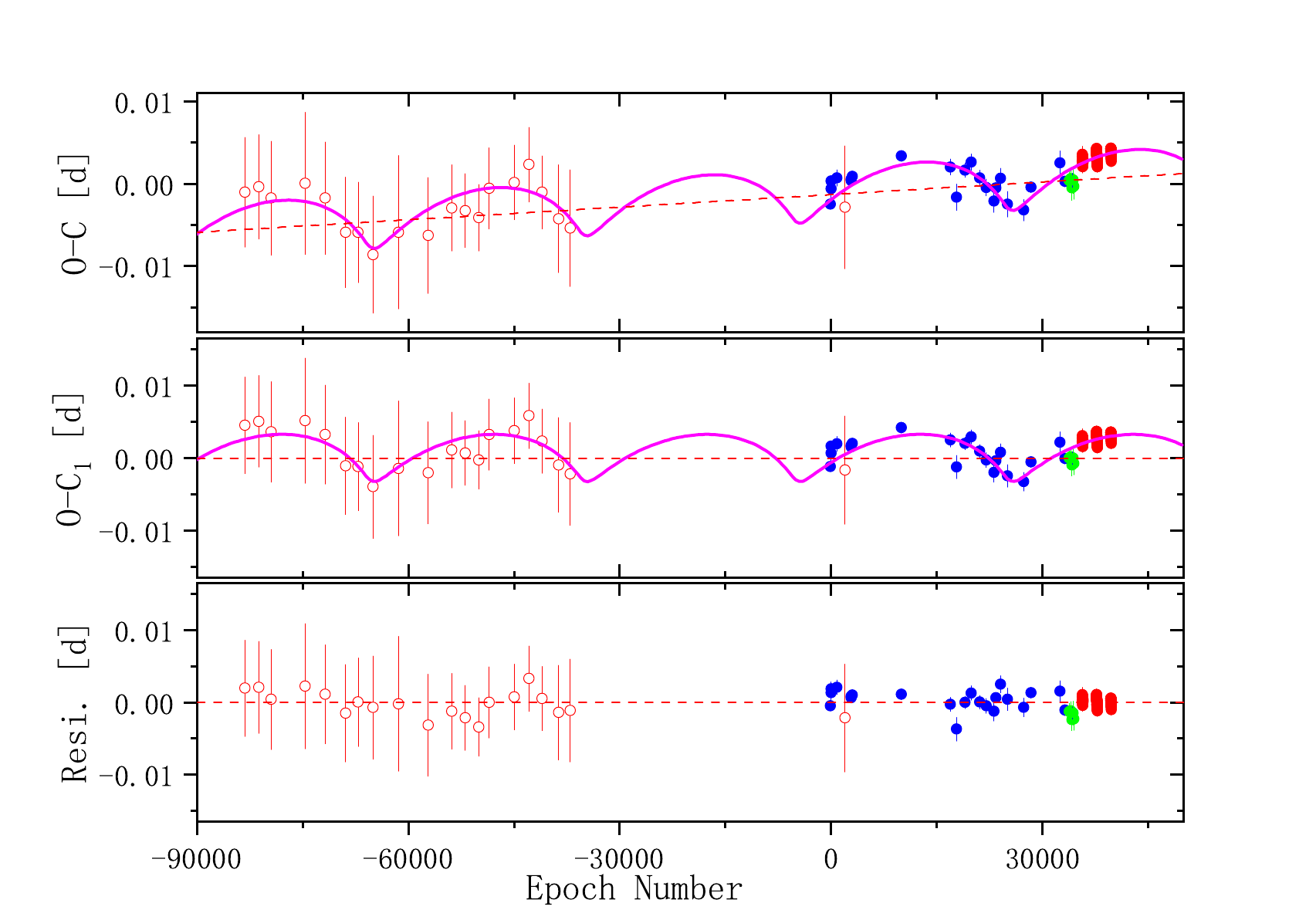}
\caption{Upper panel: the O-C diagram of KV Vel constructed with the linear ephemeris in Eq. \ref{eqn4}. The red open circles refer to the photographic (PG) data from DASCH, while red dots to those obtained from TESS. Blue and green dots represent times of light maximum collected from literature and derived from ASAS-SN. The solid magenta line suggests a combination of a revised linear ephemeris and a cyclic change. Middle panel: the $O-C_1$ curve with respect to the revised linear ephemeris. Lower panel: residuals after all changes are subtracted.} \label{fig:OC}
\end{center}
\end{figure}

More recently, 14 times of light maximum were derived by \citet{2020MNRAS.493.1197R} with light curves from the All-Sky Automated Survey (ASAS)
and ASAS-SN and with their own photometric observations. All published times of light maximum were compiled by them and they improved the accuracy of the photometric ephemeris of KV Vel as following,
\begin{equation}
\label{eqn4}
T_{max} = HJD\,2445834.52742 + 0.3571125644\times{E}.
\end{equation}
By using this linear ephemeris \citep{2020MNRAS.493.1197R}, the O-C (Observed-Calculated) curve of KV Vel is constructed with all available times of maximum light, which are listed in Table \ref{tab:max}.
The corresponding O-C plot is shown in the upper panel of Fig. \ref{fig:OC}, where the red open circles, green and red dots refer to the new determined times of light maximum by using the photographic (PG) data from DASCH, the ASAS-SN and the TESS data, respectively. Blue dots represent the data collected from the literature.

The graph illustrates a periodic variation signal in the O-C curve, with a period of 29.55 years and a semi-amplitude of 0.0034 days, which can be explained as the light travel-time effect (LTTE) via the presence of a third body \citep{1952ApJ...116..211I}.
Therefore, the solutions need a combination of a cyclic change with the eccentric orbit and a revised linear ephemeris (no linear period changes $\dot{P}$ = 0) to describe the O-C curve, namely
\begin{eqnarray}
O-C&=&\triangle T_{0}+\triangle P_{0} \times E+A'[(1-e^{2})\frac{sin(\nu + \omega)}{1+ecos\nu}+esin\omega]\nonumber\\
&=&\triangle T_{0}+\triangle P_{0} \times E+A'[\sqrt{1-e^{2}}sinE^{*}cos\omega +cosE^{*}sin\omega]\textrm{,}\label{equation02}
\end{eqnarray}
\noindent where $\Delta T_{0}$ is the correction to the initial epoch, $\Delta P_{0}$ is the correction to the initial orbital period, and E is the cycle number of light maximum. $\nu$ in equation \ref{equation02} is the true anomaly and $E^{*}$ is the eccentric anomaly. The two correlations, $M=E^{*}-esinE^{*}$ and $M=\frac{2\pi}{P_{3}}(t-T)$, were used during the solution, where $M$ is the mean anomaly and t is the time of light maximum.
The time span of the data is much longer than the determined period of the LTTE, indicating the stability and reliability of the changes in the periodic oscillation. The plot of the $O-C_1$ values with respect to the new linear ephemeris, 
\begin{equation}
\label{eqn-new}
T_{max} = HJD\,2445834.52612(30) + 0.3571126153(24)\times{E}
\end{equation}
is shown in the middle panel of Fig. \ref{fig:OC}.
The solid line in the panel is the theoretical orbit of KV Vel, which suggests a third body around the barycentre of this triple system. The explanations of all the parameters in equation \ref{equation02} and the values determined from the solution of LTTE are listed in Table \ref{tab:OP}.

\section{Physical mechanisms for the cyclic period change}

\begin{figure}
\begin{center}
\includegraphics[angle=0,scale=0.3]{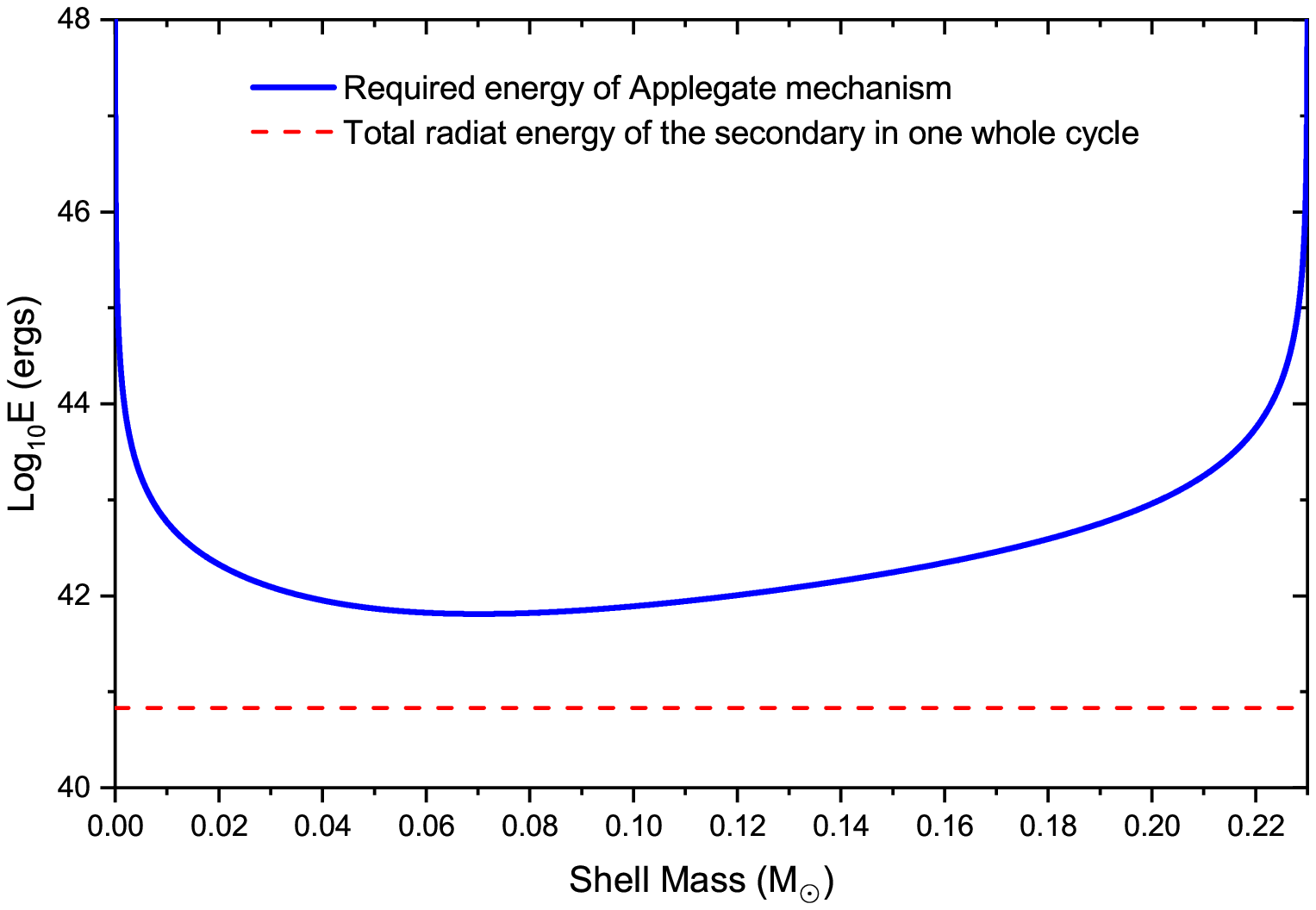}
\includegraphics[angle=0,scale=0.3]{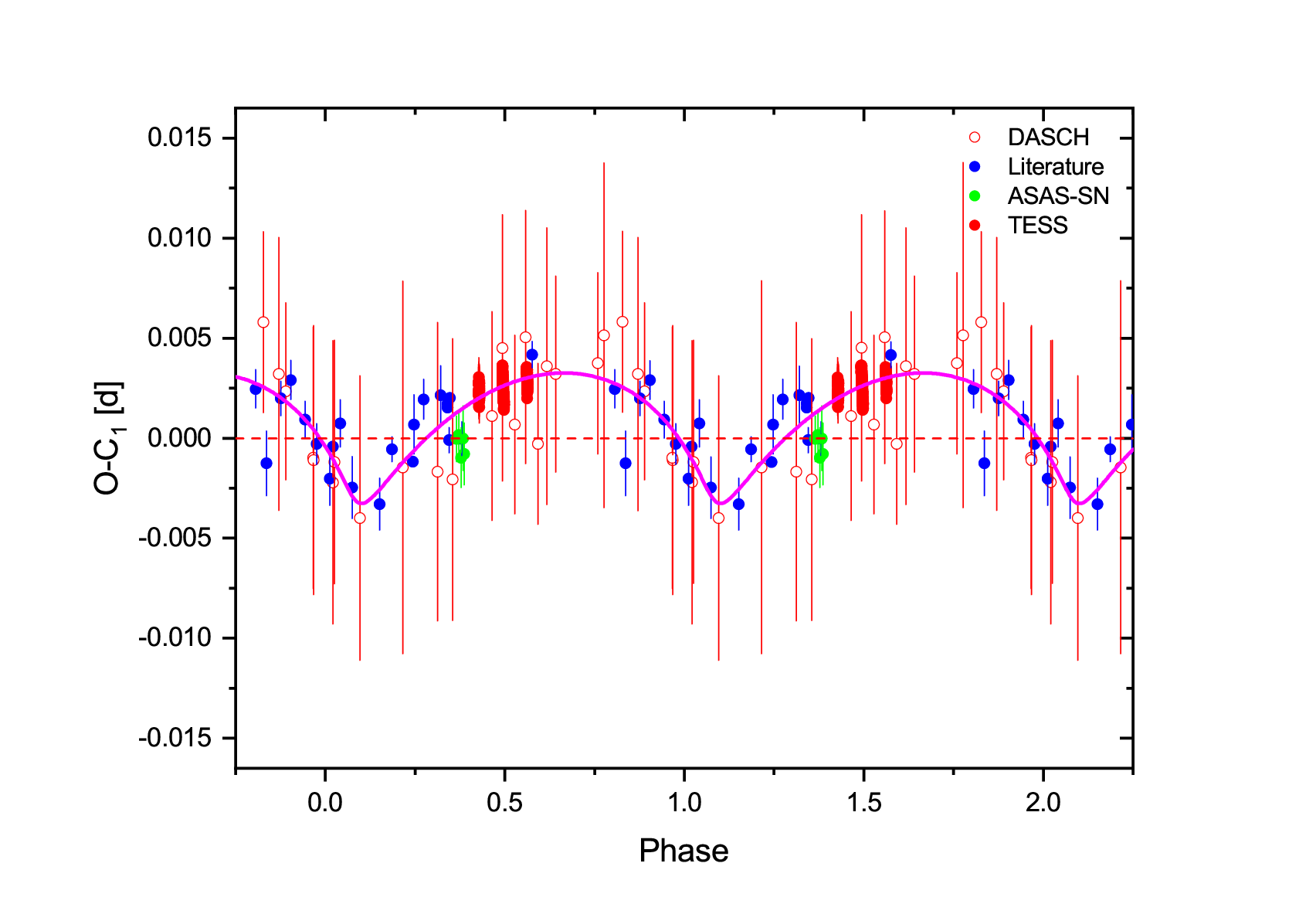}
\caption{Left panel: the required energy to produce the cyclic oscillation in
the $O-C_1$ diagram by using Applegate's mechanism (solid blue line). The red dashed line
represents the total energy that radiates from the secondary in a whole magnetic activity cycle (29.55\,years). Right panel:
the phased $O-C_1$ curve with respect to the new linear ephemeris that is caused by the presence of a low-mass companion to the central binary of the planetary nebula DS 1.} \label{fig:OC1}
\end{center}
\end{figure}

The secondary component in KV Vel is a cool M-type star with a temperature of 3400\,K and a mass of 0.23\,$M_{\odot}$. It is possible that the cyclic change in the
O-C diagram is caused by the magnetic activity cycles of
a cool component star (i.e., the Applegate mechanism) \citep{1992ApJ...385..621A}. According to this mechanism, a certain amount of angular momentum is
periodically transferred between different parts in the convection
zone of the cool star. The rotational oblateness is then change that causes the orbital period to be variable during the
cool star goes through its magnetic activity cycles. By using the same method of
\citet{2015ApJS..221...17Q}, the required energies to produce the cyclic
change in the O-C curve were computed for different shell masses of the cool component and are displayed in the left panel of Fig. \ref{fig:OC1}.
With the radius of $R_2=0.40\,R_{\odot}$ for the secondary given by \citet{1996MNRAS.279.1380H},
its luminosity was computed by using
$L_2=(\frac{R_2}{R_{\odot}})^2(\frac{T_2}{T_{\odot}})^4$\,$L_{\odot}$. The total energy radiated from the cool secondary
in a whole active cycle (29.55 years) is calculated and also displayed in the panel as the dashed line.
It is shown that the total radiated energy is much smaller than the required energies of the Applegate mechanism, which suggests
that this physical mechanism can not explain the cyclic variation of the O-C diagram.
Moreover, KV Vel was observed by TESS in four sectors from April 2019 to March 2023. The TESS light curves are symmetric and rather stable and
no signs of magnetic activities for the cool component star.

Since the Applegate mechanism has difficulty to explain the cyclic period change,
we analyzed KV Vel for the light
travel-time effect that arises from the gravitational influence of a
tertiary companion. The presence of the third body produces the
relative distance changes of the binary pair as it orbits the
barycenter of the triple system. The phased $O-C_1$ curve caused by the light
travel-time effect is displayed in the right panel of Fig. \ref{fig:OC1}. With the absolute parameters $M_1=0.63\,M_{\odot}$ and
$M_2=0.23\,M_{\odot}$ \citep{1996MNRAS.279.1380H}, a calculation by using
the following equation,
\begin{equation}
f(m)=\frac{4\pi^{2}}{GP_{3}^{2}}\times{(a_{12}^{\prime}\sin{i}^{\prime})}^{3}=\frac{(M_{3}\sin{i^{\prime}})^{3}}{(M_{1}+M_{2}+M_{3})^{2}},
\end{equation}
yields the mass function and the mass of the tertiary companion
as: $f(m)=2.6(\pm0.8)\times{10^{-4}}\,M_{\odot}$ and
$M_3\sin{i}^{\prime}=0.060(\pm0.007)\,M_{\odot}$, respectively.
$G$ in the equation is the gravitational
constant, while $P_3$ is the period of the $O-C_1$ oscillation.
The projected radius of the central binary orbiting the barycenter of the
triple system can be computed with the equation,
\begin{equation}
a_{12}^{\prime}\sin{i}^{\prime}=K\times{c}.
\end{equation}
$K$ is the amplitude of the cyclic change and $c$ is the speed of
light. When the orbital inclination of the third
body is larger than $56.5^{\circ}$, the mass of the tertiary
component corresponds to
$0.06\,M_{\odot}\le{M_3}\le0.072\,M_{\odot}$. In this case, the
tertiary component can not undergo a stable hydrogen burning in the
core, and it should be a brown dwarf. However,
depending on the unknown orbital inclination of the third body, a
low-mass stellar companion cannot be totally excluded.

\begin{table}
\begin{center}
\caption{Orbital parameters of the third body in KV Vel.} \label{tab:OP}
\begin{tabular}{lll}\hline\hline
Parameters                                      &Eccentric orbit case                 \\\hline
Revised epoch, $\Delta{T_{0}}$(d)               & $-0.0013(\pm0.0003)$               \\
Revised period, $\Delta{P_{0}}$(d)              & $5.09(\pm0.24)\times{10^{-8}}$     \\
Light travel-time effect semi-amplitude, \,A(d)                               & $0.0034(\pm0.0004)$                \\
Orbital period, $P_{3}$ (yr)                   &$29.55(\pm0.16)$                    \\
Longitude of the periastron passage, $\omega$(deg) &247.1($\pm10.2$)                 \\
Eccentricity, $e$                               & $0.67(\pm0.11)$                    \\
Projected semi-major axis, $a_{12}\sin{i_{3}}$ (au) &$0.61(\pm0.07)$            \\
$f(m)$ ($M_{\odot}$) &$0.00026(\pm0.00008)$            \\
Projected masses, $M_3\sin{i_{3}}$($M_{\odot}$) &$0.060(\pm0.007)$                \\
Maximum distances, $d_{max}$  &$9.35(\pm2.41)$  \\
\hline
\end{tabular}
\end{center}
\end{table}

\section{Discussions and conclusions}

By using several photometric databases including DASCH (the Digital Access to a Sky Century @ Harvard; \citealt{2009ASPC..410..101G,2012IAUS..285...29G}), ASAS-SN \citep{2014ApJ...788...48S,2019MNRAS.486.1907J} and TESS \citep{2015JATIS...1a4003R}, a large number of times of light maximum have been determined. The O-C diagram is constructed with these new times of maximum together with those collected from the literature. It is discovered that the O-C curve shows a cyclic oscillation with a period of 29.55\,years and an amplitude of 0.0034\,days. If the cyclic change is caused by the  magnetic activity cycles of the cool component, the required energies are much larger than the total energy radiated from the cool secondary
in a whole active cycle (29.55 years). This suggests that the cyclic variation can not be explained by the Applegate mechanism.
Therefore, we analyze the cyclic period oscillation for the light
travel-time effect that is caused by the wobble of the binary's barycentre via the existence of a third
body. The mass function and the mass of the tertiary companion are revised
as: $f(m)=2.6(\pm0.8)\times{10^{-4}}\,M_{\odot}$ and
$M_3\sin{i}^{\prime}=0.060(\pm0.007)\,M_{\odot}$, respectively.
If the orbital inclination of the third body is larger than $56.5^{\circ}$, it should be a brown dwarf and it
can not undergo stable hydrogen burning in the core. By considering that the third body is coplanar with the central binary (i.e., $i^{\prime}\ge{62.5}^{\circ}$),
the mass of the tertiary component is calculated as $M_3\sim0.068$\,$M_{\odot}$.

KV Vel is the central binary star of the planetary nebula DS 1 \citep{1983ApJ...270L..13D,1978A&AS...31...15H}.
Several investigators indicated that the initial primary in the progenitor of KV Vel filled its critical Roche lobe during the early
AGB stage (e.g., \citealt{1993ApJ...418..343I}). It had a small degenerate CO core with a helium-burning shell and an extended hydrogen-rich envelope (with initial mass in the range of 2.3-8\,$M_{\odot}$). Most of the
hydrogen-rich envelope was ejected during the common-envelope evolution, and the planetary nebula now around KV Vel is the ejected common envelope. The presence of the planetary nebula together with the extremely high temperature of the sdO primary suggests that it is one of the youngest
Post-common envelope binaries (e.g., \citealt{2007A&A...469..297A}).

When the orbital inclination equals $90^{\circ}$, the orbital radius $d_3$ of the
tertiary component in the KV Vel triple system is about 9.35\,AU, which is much smaller than the size of the roughly circular planetary nebula (with a diameter of $\sim$ $180^{\prime \prime}$). This indicates that the triple system is at the nuclei of the circular nebula. It is possible that the low-mass companion of KV Vel is formed during the ejection of the common envelope.
The loss of the AGB envelope in the equatorial plane (e.g., \citealt{1998ApJ...500..909S}), followed by a spherical fast wind from the hot primary, will cause a great quantity of mass loss. The ejected common envelope formed the planetary nebula as well as the low-mass third body.
KV Vel will evolve into normal cataclysmic variables (CV) through angular momentum loss (e.g., \citealt{2006A&A...456.1069S}). The substellar objects orbiting some cataclysmic variables, such as V2051 Oph \citep{2015ApJS..221...17Q}, DV UMa \citep{2017AJ....153..238H}, SW Sex \citep{2020ApJ...901..113F}, and HT Cas \citep{2023ApJ...953...63H}, may be also formed during the common-envelope evolution.

 This work is supported by National Key R\&D Program of China (grant No.2022YFE0116800) and National Natural
Science Foundation of China (grant Nos.11933008, 11922306, 11703083, and 11903076). The continuous photometric data used in this study
are collected by the TESS mission. Funding for the TESS mission is provided by NASA Science Mission Directorate.
We really appreciate the TESS team for supporting of this work. This work also makes use the photographic data of DASCH and the ASAS-SN database.

\appendix
\addcontentsline{toc}{section}{}
\label{sec:appendix}
In table \ref{tab:}, we present all the times of light maximum for KV Vel.

\renewcommand\thetable{A1}
\begin{tiny}
\begin{center}
\setlength{\tabcolsep}{1.1mm}{
\begin{longtable}{cccccccccccc}
\caption{\label{tab:} All the times of light maximum for KV Vel.}\\
\hline\hline
Times	&	Errors	&	Epoch	&	O-C	&	Source	&	REF.	&	Times	&	Errors	&	Epoch	&	O-C	&	Source	&	REF.	\\	
HJD\,2400000+	&	±days	&		&	days	&		&		&	HJD\,2400000+	&	±days	&		&	days	&		&		\\	
\hline
\endfirsthead 

\caption{\label{tab:}(Continued)}\\
\hline\hline
Times	&	Errors	&	Epoch	&	O-C	&	Source	&	REF.	&	Times	&	Errors	&	Epoch	&	O-C	&	Source	&	REF.	\\	
HJD\,2400000+	&	±days	&		&	days	&		&		&	HJD\,2400000+	&	±days	&		&	days	&		&		\\	

\hline \endhead 

\hline
\multicolumn{6}{r}{\textsl{(Continued)}}\\
\endfoot

\hline

\endlastfoot
16120.26124 	&	0.00666 	&	-83207	&	-0.00103	&	DASCH-PG	&	1	&	59301.24577 	&	0.00012 	&	37710	&	0.00354	&	TESS-CCD	&	1	\\
16818.05983 	&	0.00632 	&	-81253	&	-0.00039	&	DASCH-PG	&	1	&	59301.60270 	&	0.00011 	&	37711	&	0.00337	&	TESS-CCD	&	1	\\
17456.57573 	&	0.00691 	&	-79465	&	-0.00176	&	DASCH-PG	&	1	&	59301.95983 	&	0.00012 	&	37712	&	0.00338	&	TESS-CCD	&	1	\\
19171.78917 	&	0.00862 	&	-74662	&	0.00003	&	DASCH-PG	&	1	&	59302.31672 	&	0.00015 	&	37713	&	0.00316	&	TESS-CCD	&	1	\\
20187.77262 	&	0.00682 	&	-71817	&	-0.00176	&	DASCH-PG	&	1	&	59302.67403 	&	0.00013 	&	37714	&	0.00336	&	TESS-CCD	&	1	\\
21234.46540 	&	0.00672 	&	-68886	&	-0.00591	&	DASCH-PG	&	1	&	59303.03111 	&	0.00013 	&	37715	&	0.00333	&	TESS-CCD	&	1	\\
21858.69816 	&	0.00608 	&	-67138	&	-0.00591	&	DASCH-PG	&	1	&	59303.38817 	&	0.00015 	&	37716	&	0.00327	&	TESS-CCD	&	1	\\
22619.34522 	&	0.00710 	&	-65008	&	-0.00861	&	DASCH-PG	&	1	&	59303.74516 	&	0.00013 	&	37717	&	0.00315	&	TESS-CCD	&	1	\\
23908.88141 	&	0.00931 	&	-61397	&	-0.00589	&	DASCH-PG	&	1	&	59304.10240 	&	0.00012 	&	37718	&	0.00328	&	TESS-CCD	&	1	\\
25416.25316 	&	0.00704 	&	-57176	&	-0.00628	&	DASCH-PG	&	1	&	59304.45959 	&	0.00013 	&	37719	&	0.00336	&	TESS-CCD	&	1	\\
26596.51352 	&	0.00522 	&	-53871	&	-0.00294	&	DASCH-PG	&	1	&	59304.81644 	&	0.00015 	&	37720	&	0.00309	&	TESS-CCD	&	1	\\
27281.09798 	&	0.00447 	&	-51954	&	-0.00327	&	DASCH-PG	&	1	&	59305.17370 	&	0.00014 	&	37721	&	0.00324	&	TESS-CCD	&	1	\\
27972.82415 	&	0.00402 	&	-50017	&	-0.00414	&	DASCH-PG	&	1	&	59305.53092 	&	0.00015 	&	37722	&	0.00334	&	TESS-CCD	&	1	\\
28509.56790 	&	0.00490 	&	-48514	&	-0.00057	&	DASCH-PG	&	1	&	59305.88818 	&	0.00016 	&	37723	&	0.0035	&	TESS-CCD	&	1	\\
29779.81801 	&	0.00453 	&	-44957	&	0.00015	&	DASCH-PG	&	1	&	59308.74462 	&	0.00013 	&	37731	&	0.00304	&	TESS-CCD	&	1	\\
30513.68650 	&	0.00452 	&	-42902	&	0.00232	&	DASCH-PG	&	1	&	59309.10177 	&	0.00013 	&	37732	&	0.00307	&	TESS-CCD	&	1	\\
31182.55496 	&	0.00442 	&	-41029	&	-0.00106	&	DASCH-PG	&	1	&	59309.45875 	&	0.00013 	&	37733	&	0.00294	&	TESS-CCD	&	1	\\
32012.48136 	&	0.00653 	&	-38705	&	-0.00425	&	DASCH-PG	&	1	&	59309.81596 	&	0.00014 	&	37734	&	0.00303	&	TESS-CCD	&	1	\\
32607.78686 	&	0.00708 	&	-37038	&	-0.0054	&	DASCH-PG	&	1	&	59310.17324 	&	0.00012 	&	37735	&	0.0032	&	TESS-CCD	&	1	\\
45796.67100 	&		        &	-106	&	-0.00249	&	CTIO-PE	&	2	&	59310.53011 	&	0.00013 	&	37736	&	0.00296	&	TESS-CCD	&	1	\\
45834.52680 	&	0.00148 	&	0	    &	-0.00062	&	CTIO-PE	&	2	&	59310.88714 	&	0.00012 	&	37737	&	0.00288	&	TESS-CCD	&	1	\\
45835.59908 	&		        &	3	    &	0.00032	&	CTIO-PE	&	2	&	59311.24419 	&	0.00012 	&	37738	&	0.00281	&	TESS-CCD	&	1	\\
46130.57442 	&	0.00099 	&	829	&	0.00068	&	CTIO-PE	&	2	&	59311.60144 	&	0.00012 	&	37739	&	0.00295	&	TESS-CCD	&	1	\\
46538.75052 	&	0.00745 	&	1972	&	-0.00288	&	DASCH-PG	&	1	&	59311.95837 	&	0.00012 	&	37740	&	0.00277	&	TESS-CCD	&	1	\\
46846.58500 	&		&	2834	&	0.00057	&	SAAO-PM	&	3	&	59312.31565 	&	0.00013 	&	37741	&	0.00294	&	TESS-CCD	&	1	\\
46850.51304 	&		&	2845	&	0.00037	&	SAAO-PM	&	3	&	59312.67274 	&	0.00011 	&	37742	&	0.00292	&	TESS-CCD	&	1	\\
46912.29400 	&		&	3018	&	0.00086	&	SAAO-PM	&	3	&	59313.02987 	&	0.00012 	&	37743	&	0.00293	&	TESS-CCD	&	1	\\
49382.08700 	&	0.00066 	&	9934	&	0.00337	&	CCD	&	4	&	59313.38696 	&	0.00013 	&	37744	&	0.00291	&	TESS-CCD	&	1	\\
51870.08889 	&	0.00096 	&	16901	&	0.00202	&	ASAS-CCD	&	5	&	59313.74430 	&	0.00011 	&	37745	&	0.00313	&	TESS-CCD	&	1	\\
52193.98632 	&	0.00161 	&	17808	&	-0.00165	&	ASAS-CCD	&	5	&	59314.10112 	&	0.00013 	&	37746	&	0.00285	&	TESS-CCD	&	1	\\
52622.88181 	&	0.00086 	&	19009	&	0.00165	&	ASAS-CCD	&	5	&	59314.45860 	&	0.00011 	&	37747	&	0.00321	&	TESS-CCD	&	1	\\
52925.00000 	&	0.00097 	&	19855	&	0.00261	&	ASAS-CCD	&	5	&	59314.81542 	&	0.00014 	&	37748	&	0.00292	&	TESS-CCD	&	1	\\
53358.17562 	&	0.00092 	&	21068	&	0.00069	&	ASAS-CCD	&	5	&	59315.17249 	&	0.00012 	&	37749	&	0.00287	&	TESS-CCD	&	1	\\
53703.14519 	&	0.00101 	&	22034	&	-0.00047	&	ASAS-CCD	&	5	&	59315.52950 	&	0.00013 	&	37750	&	0.00277	&	TESS-CCD	&	1	\\
54090.96777 	&	0.00134 	&	23120	&	-0.00214	&	ASAS-CCD	&	5	&	59315.88668 	&	0.00012 	&	37751	&	0.00284	&	TESS-CCD	&	1	\\
54184.53287 	&		&	23382	&	-0.00053	&	NIR	&	6	&	59316.24366 	&	0.00011 	&	37752	&	0.00271	&	TESS-CCD	&	1	\\
54412.01477 	&	0.00118 	&	24019	&	0.00067	&	ASAS-CCD	&	5	&	59316.60087 	&	0.00013 	&	37753	&	0.00281	&	TESS-CCD	&	1	\\
54764.12459 	&	0.00154 	&	25005	&	-0.0025	&	ASAS-CCD	&	5	&	59316.95793 	&	0.00012 	&	37754	&	0.00275	&	TESS-CCD	&	1	\\
55595.83905 	&	0.00129 	&	27334	&	-0.00321	&	CCD	&	5	&	59317.31497 	&	0.00013 	&	37755	&	0.00268	&	TESS-CCD	&	1	\\
55971.16715 	&	0.00062 	&	28385	&	-0.00041	&	CCD	&	5	&	59317.67208 	&	0.00012 	&	37756	&	0.00268	&	TESS-CCD	&	1	\\
57423.18975 	&	0.00146 	&	32451	&	0.0025	&	ASASSN-CCD	&	5	&	59318.02923 	&	0.00012 	&	37757	&	0.00272	&	TESS-CCD	&	1	\\
57683.16548 	&	0.00063 	&	33179	&	0.00029	&	ASASSN-CCD	&	5	&	59318.38631 	&	0.00012 	&	37758	&	0.00268	&	TESS-CCD	&	1	\\
57899.21868 	&	0.00114 	&	33784	&	0.00039	&	ASASSN-CCD	&	1	&	59318.74331 	&	0.00012 	&	37759	&	0.00257	&	TESS-CCD	&	1	\\
57974.56963 	&	0.00126 	&	33995	&	0.00058	&	ASASSN-CCD	&	1	&	59321.95743 	&	0.00014 	&	37768	&	0.00268	&	TESS-CCD	&	1	\\
58042.06279 	&	0.00150 	&	34184	&	-0.00054	&	ASASSN-CCD	&	1	&	59322.31448 	&	0.00013 	&	37769	&	0.00261	&	TESS-CCD	&	1	\\
58072.06119 	&	0.00084 	&	34268	&	0.00041	&	ASASSN-CCD	&	5	&	59322.67180 	&	0.00012 	&	37770	&	0.00283	&	TESS-CCD	&	1	\\
58099.55889 	&	0.00162 	&	34345	&	0.00044	&	ASASSN-CCD	&	1	&	59323.02904 	&	0.00013 	&	37771	&	0.00295	&	TESS-CCD	&	1	\\
58135.26937 	&	0.00154 	&	34445	&	-0.00033	&	ASASSN-CCD	&	1	&	59323.38588 	&	0.00011 	&	37772	&	0.00268	&	TESS-CCD	&	1	\\
58570.94976 	&	0.00091 	&	35665	&	0.00273	&	TESS-CCD	&	1	&	59323.74290 	&	0.00012 	&	37773	&	0.00258	&	TESS-CCD	&	1	\\
58571.30742 	&	0.00091 	&	35666	&	0.00327	&	TESS-CCD	&	1	&	59324.09994 	&	0.00013 	&	37774	&	0.00251	&	TESS-CCD	&	1	\\
58571.66454 	&	0.00077 	&	35667	&	0.00328	&	TESS-CCD	&	1	&	59324.45696 	&	0.00014 	&	37775	&	0.00242	&	TESS-CCD	&	1	\\
58572.02113 	&	0.00073 	&	35668	&	0.00276	&	TESS-CCD	&	1	&	59324.81414 	&	0.00012 	&	37776	&	0.00249	&	TESS-CCD	&	1	\\
58572.37827 	&	0.00065 	&	35669	&	0.00279	&	TESS-CCD	&	1	&	59325.17121 	&	0.00013 	&	37777	&	0.00244	&	TESS-CCD	&	1	\\
58572.73615 	&	0.00098 	&	35670	&	0.00356	&	TESS-CCD	&	1	&	59325.52830 	&	0.00012 	&	37778	&	0.00242	&	TESS-CCD	&	1	\\
58573.09301 	&	0.00057 	&	35671	&	0.00331	&	TESS-CCD	&	1	&	59325.88546 	&	0.00013 	&	37779	&	0.00246	&	TESS-CCD	&	1	\\
58573.80658 	&	0.00051 	&	35673	&	0.00265	&	TESS-CCD	&	1	&	59326.24217 	&	0.00016 	&	37780	&	0.00207	&	TESS-CCD	&	1	\\
58574.16422 	&	0.00081 	&	35674	&	0.00318	&	TESS-CCD	&	1	&	59326.59972 	&	0.00013 	&	37781	&	0.00251	&	TESS-CCD	&	1	\\
58574.52103 	&	0.00075 	&	35675	&	0.00288	&	TESS-CCD	&	1	&	59326.95659 	&	0.00012 	&	37782	&	0.00226	&	TESS-CCD	&	1	\\
58574.87797 	&	0.00066 	&	35676	&	0.0027	&	TESS-CCD	&	1	&	59327.31373 	&	0.00013 	&	37783	&	0.00229	&	TESS-CCD	&	1	\\
58575.23478 	&	0.00074 	&	35677	&	0.0024	&	TESS-CCD	&	1	&	59327.67085 	&	0.00013 	&	37784	&	0.0023	&	TESS-CCD	&	1	\\
58575.59273 	&	0.00096 	&	35678	&	0.00324	&	TESS-CCD	&	1	&	59328.02810 	&	0.00012 	&	37785	&	0.00244	&	TESS-CCD	&	1	\\
58576.66293 	&	0.00062 	&	35681	&	0.0021	&	TESS-CCD	&	1	&	59328.38517 	&	0.00013 	&	37786	&	0.00239	&	TESS-CCD	&	1	\\
58577.02081 	&	0.00072 	&	35682	&	0.00287	&	TESS-CCD	&	1	&	59328.74233 	&	0.00012 	&	37787	&	0.00244	&	TESS-CCD	&	1	\\
58577.37771 	&	0.00069 	&	35683	&	0.00265	&	TESS-CCD	&	1	&	59329.09946 	&	0.00011 	&	37788	&	0.00246	&	TESS-CCD	&	1	\\
58578.09226 	&	0.00061 	&	35685	&	0.00298	&	TESS-CCD	&	1	&	59329.45656 	&	0.00013 	&	37789	&	0.00245	&	TESS-CCD	&	1	\\
58578.44919 	&	0.00085 	&	35686	&	0.0028	&	TESS-CCD	&	1	&	59329.81348 	&	0.00013 	&	37790	&	0.00225	&	TESS-CCD	&	1	\\
58579.16382 	&	0.00083 	&	35688	&	0.0032	&	TESS-CCD	&	1	&	59330.17037 	&	0.00013 	&	37791	&	0.00203	&	TESS-CCD	&	1	\\
58579.52060 	&	0.00067 	&	35689	&	0.00287	&	TESS-CCD	&	1	&	59330.52763 	&	0.00012 	&	37792	&	0.00217	&	TESS-CCD	&	1	\\
58579.87756 	&	0.00062 	&	35690	&	0.00271	&	TESS-CCD	&	1	&	59330.88487 	&	0.00013 	&	37793	&	0.00231	&	TESS-CCD	&	1	\\
58580.23426 	&	0.00063 	&	35691	&	0.0023	&	TESS-CCD	&	1	&	59331.24202 	&	0.00013 	&	37794	&	0.00234	&	TESS-CCD	&	1	\\
58580.59245 	&	0.00094 	&	35692	&	0.00338	&	TESS-CCD	&	1	&	59331.59890 	&	0.00013 	&	37795	&	0.00211	&	TESS-CCD	&	1	\\
58580.94913 	&	0.00054 	&	35693	&	0.00294	&	TESS-CCD	&	1	&	59331.95622 	&	0.00013 	&	37796	&	0.00231	&	TESS-CCD	&	1	\\
58581.66270 	&	0.00058 	&	35695	&	0.0023	&	TESS-CCD	&	1	&	59332.31340 	&	0.00013 	&	37797	&	0.00239	&	TESS-CCD	&	1	\\
58584.16268 	&	0.00088 	&	35702	&	0.00249	&	TESS-CCD	&	1	&	60015.47153 	&	0.00015 	&	39710	&	0.00418	&	TESS-CCD	&	1	\\
58584.51977 	&	0.00064 	&	35703	&	0.00247	&	TESS-CCD	&	1	&	60015.82875 	&	0.00013 	&	39711	&	0.00428	&	TESS-CCD	&	1	\\
58584.87649 	&	0.00079 	&	35704	&	0.00207	&	TESS-CCD	&	1	&	60016.18579 	&	0.00013 	&	39712	&	0.00421	&	TESS-CCD	&	1	\\
58585.23413 	&	0.00073 	&	35705	&	0.0026	&	TESS-CCD	&	1	&	60016.54294 	&	0.00012 	&	39713	&	0.00425	&	TESS-CCD	&	1	\\
58585.59170 	&	0.00099 	&	35706	&	0.00306	&	TESS-CCD	&	1	&	60016.89984 	&	0.00015 	&	39714	&	0.00404	&	TESS-CCD	&	1	\\
58586.66237 	&	0.00054 	&	35709	&	0.00239	&	TESS-CCD	&	1	&	60017.25692 	&	0.00014 	&	39715	&	0.004	&	TESS-CCD	&	1	\\
58587.02021 	&	0.00075 	&	35710	&	0.00311	&	TESS-CCD	&	1	&	60017.61426 	&	0.00015 	&	39716	&	0.00424	&	TESS-CCD	&	1	\\
58587.37680 	&	0.00071 	&	35711	&	0.00259	&	TESS-CCD	&	1	&	60017.97136 	&	0.00014 	&	39717	&	0.00422	&	TESS-CCD	&	1	\\
58588.09128 	&	0.00065 	&	35713	&	0.00285	&	TESS-CCD	&	1	&	60018.32845 	&	0.00013 	&	39718	&	0.00419	&	TESS-CCD	&	1	\\
58588.44849 	&	0.00095 	&	35714	&	0.00294	&	TESS-CCD	&	1	&	60018.68553 	&	0.00012 	&	39719	&	0.00417	&	TESS-CCD	&	1	\\
58589.16304 	&	0.00088 	&	35716	&	0.00327	&	TESS-CCD	&	1	&	60019.04272 	&	0.00013 	&	39720	&	0.00424	&	TESS-CCD	&	1	\\
58589.52005 	&	0.00073 	&	35717	&	0.00316	&	TESS-CCD	&	1	&	60019.39979 	&	0.00013 	&	39721	&	0.0042	&	TESS-CCD	&	1	\\
58589.87678 	&	0.00071 	&	35718	&	0.00278	&	TESS-CCD	&	1	&	60019.75651 	&	0.00013 	&	39722	&	0.00381	&	TESS-CCD	&	1	\\
58590.23364 	&	0.00066 	&	35719	&	0.00253	&	TESS-CCD	&	1	&	60020.11371 	&	0.00013 	&	39723	&	0.00389	&	TESS-CCD	&	1	\\
58590.94826 	&	0.00064 	&	35721	&	0.00292	&	TESS-CCD	&	1	&	60020.47089 	&	0.00014 	&	39724	&	0.00396	&	TESS-CCD	&	1	\\
58591.66196 	&	0.00051 	&	35723	&	0.00241	&	TESS-CCD	&	1	&	60020.82804 	&	0.00013 	&	39725	&	0.004	&	TESS-CCD	&	1	\\
58592.01968 	&	0.00076 	&	35724	&	0.00301	&	TESS-CCD	&	1	&	60021.54222 	&	0.00011 	&	39727	&	0.00396	&	TESS-CCD	&	1	\\
58592.37653 	&	0.00076 	&	35725	&	0.00275	&	TESS-CCD	&	1	&	60021.89961 	&	0.00013 	&	39728	&	0.00423	&	TESS-CCD	&	1	\\
58593.09139 	&	0.00059 	&	35727	&	0.00338	&	TESS-CCD	&	1	&	60022.25648 	&	0.00012 	&	39729	&	0.00399	&	TESS-CCD	&	1	\\
58593.44812 	&	0.00092 	&	35728	&	0.003	&	TESS-CCD	&	1	&	60022.61333 	&	0.00012 	&	39730	&	0.00373	&	TESS-CCD	&	1	\\
58594.16264 	&	0.00092 	&	35730	&	0.0033	&	TESS-CCD	&	1	&	60022.97078 	&	0.00014 	&	39731	&	0.00406	&	TESS-CCD	&	1	\\
58594.51970 	&	0.00068 	&	35731	&	0.00324	&	TESS-CCD	&	1	&	60023.32784 	&	0.00013 	&	39732	&	0.00401	&	TESS-CCD	&	1	\\
58594.87645 	&	0.00071 	&	35732	&	0.00288	&	TESS-CCD	&	1	&	60023.68495 	&	0.00013 	&	39733	&	0.00401	&	TESS-CCD	&	1	\\
58595.23327 	&	0.00064 	&	35733	&	0.00259	&	TESS-CCD	&	1	&	60024.04206 	&	0.00012 	&	39734	&	0.004	&	TESS-CCD	&	1	\\
59282.31918 	&	0.00013 	&	37657	&	0.00392	&	TESS-CCD	&	1	&	60024.39931 	&	0.00014 	&	39735	&	0.00415	&	TESS-CCD	&	1	\\
59282.67652 	&	0.00012 	&	37658	&	0.00415	&	TESS-CCD	&	1	&	60024.75641 	&	0.00010 	&	39736	&	0.00413	&	TESS-CCD	&	1	\\
59283.03340 	&	0.00014 	&	37659	&	0.00392	&	TESS-CCD	&	1	&	60025.11341 	&	0.00014 	&	39737	&	0.00402	&	TESS-CCD	&	1	\\
59283.39082 	&	0.00013 	&	37660	&	0.00422	&	TESS-CCD	&	1	&	60025.47043 	&	0.00014 	&	39738	&	0.00393	&	TESS-CCD	&	1	\\
59283.74796 	&	0.00013 	&	37661	&	0.00425	&	TESS-CCD	&	1	&	60025.82727 	&	0.00013 	&	39739	&	0.00365	&	TESS-CCD	&	1	\\
59284.10491 	&	0.00013 	&	37662	&	0.00409	&	TESS-CCD	&	1	&	60026.18462 	&	0.00012 	&	39740	&	0.00389	&	TESS-CCD	&	1	\\
59284.46213 	&	0.00013 	&	37663	&	0.0042	&	TESS-CCD	&	1	&	60026.54173 	&	0.00015 	&	39741	&	0.00389	&	TESS-CCD	&	1	\\
59284.81927 	&	0.00013 	&	37664	&	0.00422	&	TESS-CCD	&	1	&	60026.89857 	&	0.00015 	&	39742	&	0.00362	&	TESS-CCD	&	1	\\
59285.17632 	&	0.00012 	&	37665	&	0.00416	&	TESS-CCD	&	1	&	60027.25593 	&	0.00013 	&	39743	&	0.00386	&	TESS-CCD	&	1	\\
59285.53324 	&	0.00014 	&	37666	&	0.00397	&	TESS-CCD	&	1	&	60027.96978 	&	0.00014 	&	39745	&	0.00349	&	TESS-CCD	&	1	\\
59285.89019 	&	0.00012 	&	37667	&	0.00381	&	TESS-CCD	&	1	&	60028.32731 	&	0.00013 	&	39746	&	0.0039	&	TESS-CCD	&	1	\\
59286.24737 	&	0.00014 	&	37668	&	0.00387	&	TESS-CCD	&	1	&	60028.68424 	&	0.00014 	&	39747	&	0.00372	&	TESS-CCD	&	1	\\
59286.60454 	&	0.00014 	&	37669	&	0.00394	&	TESS-CCD	&	1	&	60029.04120 	&	0.00015 	&	39748	&	0.00357	&	TESS-CCD	&	1	\\
59286.96172 	&	0.00014 	&	37670	&	0.004	&	TESS-CCD	&	1	&	60029.39846 	&	0.00013 	&	39749	&	0.00372	&	TESS-CCD	&	1	\\
59287.31909 	&	0.00012 	&	37671	&	0.00425	&	TESS-CCD	&	1	&	60029.75565 	&	0.00012 	&	39750	&	0.00379	&	TESS-CCD	&	1	\\
59287.67604 	&	0.00013 	&	37672	&	0.0041	&	TESS-CCD	&	1	&	60030.11274 	&	0.00013 	&	39751	&	0.00378	&	TESS-CCD	&	1	\\
59288.03289 	&	0.00013 	&	37673	&	0.00383	&	TESS-CCD	&	1	&	60030.46995 	&	0.00015 	&	39752	&	0.00387	&	TESS-CCD	&	1	\\
59288.39009 	&	0.00013 	&	37674	&	0.00392	&	TESS-CCD	&	1	&	60030.82697 	&	0.00011 	&	39753	&	0.00378	&	TESS-CCD	&	1	\\
59288.74706 	&	0.00013 	&	37675	&	0.00378	&	TESS-CCD	&	1	&	60031.18405 	&	0.00013 	&	39754	&	0.00374	&	TESS-CCD	&	1	\\
59289.10426 	&	0.00012 	&	37676	&	0.00386	&	TESS-CCD	&	1	&	60031.54117 	&	0.00012 	&	39755	&	0.00376	&	TESS-CCD	&	1	\\
59289.46115 	&	0.00011 	&	37677	&	0.00364	&	TESS-CCD	&	1	&	60031.89821 	&	0.00014 	&	39756	&	0.00368	&	TESS-CCD	&	1	\\
59289.81863 	&	0.00014 	&	37678	&	0.00401	&	TESS-CCD	&	1	&	60032.25507 	&	0.00013 	&	39757	&	0.00343	&	TESS-CCD	&	1	\\
59290.17523 	&	0.00013 	&	37679	&	0.0035	&	TESS-CCD	&	1	&	60032.61195 	&	0.00013 	&	39758	&	0.00319	&	TESS-CCD	&	1	\\
59290.53273 	&	0.00012 	&	37680	&	0.00388	&	TESS-CCD	&	1	&	60032.96934 	&	0.00013 	&	39759	&	0.00347	&	TESS-CCD	&	1	\\
59290.88988 	&	0.00014 	&	37681	&	0.00392	&	TESS-CCD	&	1	&	60033.32650 	&	0.00013 	&	39760	&	0.00352	&	TESS-CCD	&	1	\\
59291.24670 	&	0.00014 	&	37682	&	0.00363	&	TESS-CCD	&	1	&	60033.68306 	&	0.00014 	&	39761	&	0.00296	&	TESS-CCD	&	1	\\
59291.60400 	&	0.00012 	&	37683	&	0.00381	&	TESS-CCD	&	1	&	60034.03992 	&	0.00019 	&	39762	&	0.00271	&	TESS-CCD	&	1	\\
59291.96111 	&	0.00014 	&	37684	&	0.00381	&	TESS-CCD	&	1	&	60034.39748 	&	0.00014 	&	39763	&	0.00316	&	TESS-CCD	&	1	\\
59292.31813 	&	0.00013 	&	37685	&	0.00372	&	TESS-CCD	&	1	&	60034.75483 	&	0.00013 	&	39764	&	0.0034	&	TESS-CCD	&	1	\\
59292.67544 	&	0.00013 	&	37686	&	0.00392	&	TESS-CCD	&	1	&	60035.11204 	&	0.00013 	&	39765	&	0.00349	&	TESS-CCD	&	1	\\
59295.53198 	&	0.00014 	&	37694	&	0.00356	&	TESS-CCD	&	1	&	60035.46916 	&	0.00013 	&	39766	&	0.0035	&	TESS-CCD	&	1	\\
59295.88891 	&	0.00013 	&	37695	&	0.00337	&	TESS-CCD	&	1	&	60035.82604 	&	0.00012 	&	39767	&	0.00327	&	TESS-CCD	&	1	\\
59296.24633 	&	0.00012 	&	37696	&	0.00368	&	TESS-CCD	&	1	&	60036.18356 	&	0.00013 	&	39768	&	0.00368	&	TESS-CCD	&	1	\\
59296.60352 	&	0.00012 	&	37697	&	0.00376	&	TESS-CCD	&	1	&	60036.54045 	&	0.00015 	&	39769	&	0.00345	&	TESS-CCD	&	1	\\
59296.96045 	&	0.00013 	&	37698	&	0.00358	&	TESS-CCD	&	1	&	60036.89757 	&	0.00014 	&	39770	&	0.00346	&	TESS-CCD	&	1	\\
59297.31768 	&	0.00013 	&	37699	&	0.00369	&	TESS-CCD	&	1	&	60037.25444 	&	0.00012 	&	39771	&	0.00322	&	TESS-CCD	&	1	\\
59297.67464 	&	0.00013 	&	37700	&	0.00355	&	TESS-CCD	&	1	&	60037.61189 	&	0.00013 	&	39772	&	0.00356	&	TESS-CCD	&	1	\\
59298.03180 	&	0.00013 	&	37701	&	0.00359	&	TESS-CCD	&	1	&	60037.96894 	&	0.00013 	&	39773	&	0.0035	&	TESS-CCD	&	1	\\
59298.38896 	&	0.00013 	&	37702	&	0.00363	&	TESS-CCD	&	1	&	60038.32594 	&	0.00012 	&	39774	&	0.00338	&	TESS-CCD	&	1	\\
59298.74611 	&	0.00013 	&	37703	&	0.00368	&	TESS-CCD	&	1	&	60038.68287 	&	0.00014 	&	39775	&	0.00321	&	TESS-CCD	&	1	\\
59299.10273 	&	0.00014 	&	37704	&	0.00318	&	TESS-CCD	&	1	&	60039.04006 	&	0.00013 	&	39776	&	0.00328	&	TESS-CCD	&	1	\\
59299.45992 	&	0.00015 	&	37705	&	0.00326	&	TESS-CCD	&	1	&	60039.39743 	&	0.00013 	&	39777	&	0.00354	&	TESS-CCD	&	1	\\
59299.81715 	&	0.00012 	&	37706	&	0.00337	&	TESS-CCD	&	1	&	60039.75418 	&	0.00015 	&	39778	&	0.00317	&	TESS-CCD	&	1	\\
59300.17410 	&	0.00012 	&	37707	&	0.00322	&	TESS-CCD	&	1	&	60040.11120 	&	0.00020 	&	39779	&	0.00308	&	TESS-CCD	&	1	\\
59300.53149 	&	0.00012 	&	37708	&	0.00349	&	TESS-CCD	&	1	&	60040.46828 	&	0.00029 	&	39780	&	0.00305	&	TESS-CCD	&	1	\\
59300.88842 	&	0.00012 	&	37709	&	0.00331	&	TESS-CCD	&	1	\\												
											
\hline
\end{longtable}}
{\footnotesize References:} \footnotesize (1) This work; (2) \citet{1986AJ.....91.1372L}; (3) \citet{1988Obs...108...88K}; (4) \citet{1996MNRAS.279.1380H}; (5) \citet{2020MNRAS.493.1197R}; (6) \citet{2011A&A...526A.150R}.\\
\label{tab:max}

\end{center}
\end{tiny}

\bibliography{sample631}{}
\bibliographystyle{aasjournal}



\end{document}